\headline={\ifnum\pageno>1 \hss \number\pageno\ \hss \else\hfill \fi}

\pageno=1
\nopagenumbers
\hbadness=100000
\vbadness=100000

\centerline{\bf $A_N$ MULTIPLICITY RULES AND SCHUR FUNCTIONS}
\vskip 15mm
\centerline{\bf H. R. Karadayi \footnote{*}{
e-mail: karadayi@itu.edu.tr} }
\centerline{Dept.Physics, Fac. Science, Tech.Univ.Istanbul }
\centerline{ 80626, Maslak, Istanbul, Turkey }
\vskip 10mm
\centerline{\bf{Abstract}}
\vskip 10mm

It is well-known that explicit calculation of the right-hand side of Weyl
character formula could often be problematic in practice. We show that
a specialization in this formula can be carried out in such a way that its
right-hand side becomes simply a Schur Function. For this, we need the use
of some properly chosen set of weights which we call {\bf fundamental weights}
by the aid of which elements of Weyl orbits are expressed conveniently.

In the generic definition, an Elementary Schur Function
$S_Q(x_1,x_2,..,x_Q)$ of degree Q is known to be defined by some polynomial
of Q indeterminates $ x_1,x_2,..,x_Q $. It is also known that definition of
Elementary Schur Functions can be generalized in such a way that for any
partition $(Q_k)$ of weight Q and length k one has a Generalized Schur
Function $S_{(Q_k)}(x_1,x_2,..,x_Q)$. When they are considered for $A_{N-1}$
Lie algebras, on the other hand, a kind of degeneration occurs for these
generic definitions. This is mainly due to the fact that, for an $A_{N-1}$
Lie algebra, only a finite number of indeterminates, namely (N-1), can be
independent. This leads us to define {\bf Degenerated Schur Functions}
by taking, for $Q > N-1$, all the indeterminates $x_Q$ to be non-linearly
dependent on first (N-1) indeterminates $x_1,x_2,..,x_{N-1}$. With this in
mind, we show that for each and every dominant weight of $A_{N-1}$ we always
have a (Degenerated) Schur Function which provides the right-hand side of Weyl
character formula for irreducible representation corresponding to this dominant
weight. For instance, an Elementary Schur Function $S_Q(x_1,x_2,..,x_Q)$ is,
for $ Q < N $, nothing but the character of tensorial representation with Q
symmetrical indices for $A_{N-1}$. The same is also true for $Q \geq N$ but
there corresponds a Degenerated Elementary Schur Function.

Generalized Schur Functions are known to be expressed by determinants of some
matrices of Elementary Schur Functions. We would like to call these expressions
{\bf multiplicity rules}. This is mainly due to the fact that, to calculate
weight multiplicities, these rules give us an efficient method which works
equally well no matter how big is the rank of algebras or the dimensions of
representations.

\hfill\eject

\vskip 3mm
\noindent {\bf{I.\ INTRODUCTION}}
\vskip 3mm

The calculation of weight multiplicities is essential in many problems
encountered in group theory applications and also in theoretical high energy
physics. Due to seminal works of Freudenthal {\bf [1]}, Racah {\bf [2]} and
Kostant {\bf [3]} the problem seems to be solved to great extent for
finite Lie algebras. One must however note that great difficulties arise
in practical applications
\footnote{(*)}{ A.U.Klymik has pointed out by e-mail that the use of
Gelfand-Zeitlin basis proves also useful in calculations of weight
multiplicities }
of these multiplicity formulas when the rank of
algebras and also the dimensions of representations grow high. There are
therefore quite many works{\bf [4]} trying to reconsider the problem.

Beside these formulas, one can say that Weyl character formula {\bf [5]}
gives a unified framework to calculate the weight multiplicities for finite
and also infinite dimensional Lie algebras {\bf [6]}. As well as for finite
Lie algebras {\bf [7]} there are numerous applications of Weyl-Kac character
formula {\bf [8]} for affine Lie algebras {\bf [9]} though quite a little is
known beyond affine Lie algebras. Both for finite or affine Lie algebras,
Weyl-Kac character formula involve sums over Weyl groups of finite Lie
algebras {\bf [10]}. As we will show in a subsequent paper, the sum over the
Weyl group of a finite Lie algebra of rank N can always be cast into a sum
over the Weyl group of an $A_{N-1}$ algebra. One can thus conclude that the
weight multiplicities for any finite Lie algebra are known by calculating only
$A_N$ weight multiplicities. Therefore, it will be suitable to consider the
problem first for $A_N$ Lie algebras.

We want to show here that there is an intriguing interplay between
$A_N$ weight multiplicities and Schur functions which arise in the litterature
in a different context. Within this framework, the definition of Schur
functions plays in fact the role of a kind of multiplicity formula by giving
the right hand side of the Weyl character formula directly.

For any degree Q=1,2,... , Elementary Schur Functions $S_Q(x_1,x_2,.., x_Q)$
are defined {\bf [11]} by the formula
$$ \sum_{k \in Z} \ S_k(x_1,x_2,.., x_k) \  z^k \equiv Exp\sum_{i=1}^\infty
\ x_i \ z^i \eqno(I.1) $$
with the understanding that $S_0=1$ and also $S_{-Q} \equiv 0$.
Note here that this definition includes only some restricted class of Schur
functions of degree Q and which are polynomials of Q indeterminates
$x_1,x_2,.. x_Q$. Let us define, on the other hand, a partition $(Q_k)$ with
weight Q and length k  (=1,2,..,Q) as in the following:
$$ q_1 + q_2 + .. + q_k \equiv Q  \ \ , \ \
q_1 \geq q_2 \geq ... q_k > 0  \eqno(I.2)  $$
For any such partition $(Q_k)$, another generalization
$$ S_{(Q_k)}(x_1,x_2,..,x_Q) $$
of Schur functions are known to be defined by determinant of the following
$k \times k$ matrix:
$$ \pmatrix{
S_{q_1-0}&S_{q_1+1}&S_{q_1+2}&\ldots&S_{q_1+k-1}\cr
S_{q_2-1}&S_{q_2+0}&S_{q_2+1}&\ldots&S_{q_2+k-2}\cr
S_{q_3-2}&S_{q_3-1}&S_{q_3+0}&\ldots&S_{q_3+k-3}\cr
\vdots&\vdots&\vdots&\vdots\cr} \eqno(I.3)  $$
and hence it is natural to call them {\bf Generalized Schur Functions}.
By decomposing the determinant of matrix (I.3) in terms of its minors, one
obtains an equivalent expression for $S_{(Q_k)}(x_1,x_2,..,x_N)$
in terms of products of Elementary Schur functions. We will call these
decompositions {\bf the multiplicity rules} because, with a suitable
generalization, it will be seen in the following that they are
nothing but the right hand side of the Weyl character formula and hence
provide a powerful method in calculating weight multiplicities.
By experience, anyone can easily see in fact that this will give us a quite
fast method to calculate weight multiplicities in comparison with all the
known methods. This will be exemplified in the last section.

\vskip 3mm
\noindent {\bf{II.\ CHARACTERS OF $ A_{N-1} $ WEYL ORBITS}}
\vskip 3mm

As it is known, characters are conventionally defined for irreducible
representations of finite Lie algebras and also a huge class of representations
of affine Lie algebras. Now, rather than representations, we give a convenient
extension of the concept for Weyl orbits. It will be seen that, with
an appropriate specialization in Weyl character formula, this allows us to
recognize generalized Schur functions $S_{Q_k}(x_1,x_2,..,x_{N-1})$ directly
as $A_{N-1}$ characters.

Here, it is essential to use {\bf fundamental weights} $\mu_I$ (I=1,2,.. N)
which are defined, for $A_{N-1}$, by
$$ \eqalign {
\mu_1 &\equiv \lambda_1 \cr
\mu_i &\equiv \mu_{i-1}-\alpha_{i-1} \ \ , \ \ i=2,3,.. N. } \eqno(II.1) $$
\noindent or conversely by
$$ \lambda_i = \mu_1 + \mu_2 + .. + \mu_i \ \ , \ \ i=1,2,..N-1. \eqno(II.2) $$
\noindent together with the condition that
$$ \mu_1 + \mu_2 + .. + \mu_N \equiv 0 . \ \ \eqno(II.3)   $$
$\lambda_i$'s and $\alpha_i$'s (i=1,2,..N-1) are {\bf fundamental dominant
weights} and {\bf simple roots} of $A_{N-1}$ Lie algebras. For an excellent
study of Lie algebra technology we refer to the book of Humphreys {\bf [13]}.
We know that there is an irreducible $A_{N-1}$ representation for
each and every dominant weight $\Lambda^+$ which can be expressed by
$$ \Lambda^+ = \sum_{i=1}^K q_i \ \mu_i \ \ , \ \
q_1 \geq q_2 \geq .. \geq q_k > 0  \ \   \eqno(II.4) $$
providing in general that $k \leq N-1$. Its {\bf height} will also be defined by
$$ h(\Lambda^+) \equiv \sum_{i=1}^k \ q_i \ \ . \eqno(II.5)  $$
One thus concludes that there is a dominant weight $\Lambda^+$ for each
and every partition $(Q_k)$ with weight \break
$Q=h(\Lambda^+)$. Note here that there could be in general more than one
dominant weight with the same length. For $A_{N-1}$ Lie algebras, we also
define $Sub(Q \ \lambda_1)$ to be the set of all dominant weights
corresponding, for k=1,2,..Q, to all the partitions $(Q_k)$ with,
{\bf modulo N}, the same weigth $Q=h(\Lambda^+)$.

The weight structure of the corresponding irreducible representation
$R(\Lambda^+)$ can then be expressed by the aid of the following
{\bf orbital decomposition}:
$$ R(\Lambda^+) = \sum_{\lambda^+ \in Sub(Q \ \lambda_1)} \
m_{\Lambda^+}(\lambda^+) \ W(\lambda^+) \ \ \eqno(II.6) $$
where $m_{\Lambda^+}(\lambda^+)$ is the multiplicity of $\lambda^+$ within
$R(\Lambda^+)$. Note here that one always has
$m_{\Lambda^+}(\Lambda^+) = 1$ and also
$$m_{\Lambda^+}(\lambda^+) = 0 \ \ , \ \ K \geq k  $$
if $\Lambda^+$ and $ \lambda^+$ correspond, respectively, to partitions
$(Q_K)$ and $(Q_k)$. For instance, the tensor
representations with Q completely symmetric and antisymmetric indices
correspond, respectively, to two extreme cases which come from the partitions
$(Q_1)$ and $(Q_Q)$ provided that
$$ m_{Q \ \lambda_1}(\lambda^+) = 1 \eqno(II.7)  $$
and
$$ m_{\lambda_Q}(\lambda^+) = 0 \ \ ,
\ \ \lambda^+ \neq \lambda_Q  \eqno(II.8) $$
for any $\lambda^+ \in Sub(Q \ \lambda_1)$.

The left-hand side of Weyl character formula can now be expressed by
$$ ChR(\Lambda^+) = \sum_{\lambda^+ \in Sub(Q \ \lambda_1)} \ m_{\Lambda^+}(\lambda^+) \
ChW(\lambda^+) \ \ \eqno(II.9) $$
In view of orbital decomposition (II.6), it is clear that (II.9) allows us to
define the characters
$$ ChW(\Lambda^+) \equiv \sum_{\mu \in W(\Lambda^+)} \ e(\mu) \eqno(II.10) $$
\noindent for Weyl orbits $W(\Lambda^+)$. \ {\bf Formal exponentials} $e(\mu)$
are defined, for any weight $\mu$, as in the book of Kac {\bf [8]}.

For any partition $(Q_k)$ with weight Q and length k, it will be useful
to introduce here the class functions $K_{(Q_k)}(u_1,u_2,..,u_M)$ which are
defined to be polynomials, say, of N indeterminates $u_1,u_2,..,u_N$ as in the
following:
$$ K_{(Q_k)}(u_1,u_2,..,u_N) \equiv \sum_{j_1,j_2, .. j_k=1}^N
(u_{j_1})^{q_1} (u_{j_2})^{q_2} ...\ (u_{j_k})^{q_k} \ \ . \eqno(II.11) $$
In (II.11), no two of indices $ j_1,j_2, .. j_k $ shall take the
same value for each particular monomial. Of particular importance, one must
note also that, on the contrary to generic definitions of Elementary or
Generalized Schur Functions, the values of Q and N need not be correlated
here.

Now, to be the main result of this section, if $\Lambda^+$ corresponds
to a partition $(Q_k)$, we can state, for $A_{N-1}$, that
$$ ChW(\Lambda^+) = K_{(Q_k)}(u_1,u_2,..,u_N) \eqno(II.12) $$
in the specialization
$$ e(\mu_i) \equiv u_i \ \ , \ \ i=1,2,...N. \eqno(II.13)  $$
of (II.10). One knows here that indeterminates $u_1,u_2,..,u_N$ are restricted,
due to (II.3), by the condition that
$$ u_1 \ u_2 \ ... \ u_N \equiv 1 \ \ .  \eqno(II.14) $$

In order to investigate (II.12), we refer to a permutational lemma which we
introduced quite previously {\bf[12]}. In view of this lemma, for any dominant
weight $\Lambda^+$ which corresponds to a partition $(Q_k)$, all other weights
of the Weyl orbit $W(\Lambda^+)$ are to be obtained by all permutations of
parameters $q_1,q_2,..,q_k$.

\vskip 3mm
\noindent {\bf{III.\ REDUCTION RULES AND SCHUR FUNCTIONS}}
\vskip 3mm

In this section, we want to expose the explicit correspondence which governs
on one side $A_{N-1}$ characters defined as in (II.9) and (Generalized) Schur
Functions $S_{(Q_k)}(x_1,x_2,..,x_{N-1})$ on the other side.
For this, we first emphasize that the set of indeterminates $x_1,x_2,..$ are
conceptually different from indeterminates $u_1,u_2,..$ which give us the
specialization (II.13) in Weyl character formula. Within the framework
of our work, we however suggest a correspondence determined by the replacements
$$ K(Q) \rightarrow Q \ x_Q  \ \ , \ \  Q=1,2,...   \eqno(III.1) $$
In the notation of (II.11), the generators K(Q) are defined,
for partition $(Q_1)$, as
$$ K(Q) \equiv K_{(Q_1)}(u_1,u_2,..,u_N) \ \ . \eqno(III.2) $$
The reason why we call K(Q)'s generators is the fact that they generate
all other class functions \break
$K_{(Q_k)}(u_1,u_2,..,u_N)$ by the aid  of some
{\bf reduction rules}. By suppressing explicit $u_i$ dependences, these
reduction rules are given by
$$ \eqalign {
&K_{(q_1,q_2)} = K(q_1) \ K(q_2) - K(q_1+q_2) \ \ , \ \ q_1 > q_2 \cr
&K_{(q_1,q_1)} = {1 \over 2} K(q_1) \ K(q_1) - {1 \over 2} K(q_1+q_1)  \cr
&K_{(q_1,q_2,q_3)} = K(q_1) \ K_{(q_2,q_3)} - K_{(q_1+q_2,q_3)} - K_{(q_1+q_3,q_2)} \ \ , \ \ q_1 > q_2 > q_3  \cr
&K_{(q_1,q_2,q_2)} = K(q_1) \ K_{(q_2,q_2)} - K_{(q_1+q_2,q_2)} \ \ , \ \ q_1 > q_2 \cr
&K_{(q_1,q_1,q_2)} = {1 \over 2} K(q_1) \ K_{(q_1,q_2)} - {1 \over 2} K_{(q_1+q_1,q_2)} -
{1 \over 2} K_{(q_1+q_2,q_1)} \ \ , \ \ q_1 > q_2  \cr
&K_{(q_1,q_1,q_1)} = {1 \over 3} K(q_1) \ K_{(q_1,q_1)} - {1 \over 3} K_{(q_1+q_1,q_1)} } \eqno(III.3) $$
for the first three orders. For higher orders, the reduction
rules can be obtained similarly as in above. This allows us to express
everything in terms of generators K(Q) and hence the name generators.

In view of (II.9) and (II.7), one obtains for any given $A_{N-1}$
$$ ChR(Q \ \lambda_1) = \sum \ K_{(Q_k)}(u_1,u_2,..,u_N) \eqno(III.4) $$
where the sum is over all partitions $(Q_k)$ with weight Q and length
k=1,2,..Q. For the right-hand side of (III.4), an equivalent expression can be
provided here by polynomials $h_k(u_1,u_2,..,u_k)$ defined in
$$ \sum_{k=0}^\infty \ h_k(u_1,u_2,..,u_k) \ z^k \equiv
\prod_{i=1}^\infty \ {1 \over (1-z \ u_i)} \ \ . \eqno(III.5) $$
By applying reduction rules mentioned above, one clearly sees, in view of
(III.1), the equivalence
$$ h_Q(u_1,u_2,..,u_Q) \rightarrow S_Q(x_1,x_2,..,x_Q) $$
and hence one obtains
$$ ChR(Q \ \lambda_1) \equiv S_Q(x_1,x_2,..,x_Q) \eqno(III.6) $$

Now, it will be instructive to give here some clarifications on how we use
Schur functions in applications of Weyl character formula for $A_{N-1}$
Lie algebras. For this, we first note that in their generic definitions (I.1)
there is no reference to any Lie algebra and hence $S_Q(x_1,x_2,..,x_{N-1})$
can only be defined providing Q = N-1, i.e. a Schur function of degree Q can
only be expressed in terms of Q indeterminates.
For the reader's convenience, we called these {\bf Elementary Schur Functions}.
In our applications for $A_{N-1}$, due to influence of condition (II.3),
we are in the situation to introduce the concept of {\bf Degenerated Schur
Functions}. As the first example, one can say that
$S_Q(x_1,x_2,..,x_{N-1})$ is degenerated for Q = N in the sense that if it
is represented by a polinomial for $A_{N-1}$ it will be represented by a
different one for all other Lie algebras $A_N,A_{N+1},..$. To this end,
let us consider
$$ K_{(N_N)}(u_1,u_2,..,u_N) \equiv 1 \ \ .  \eqno(III.7) $$
which is nothing but the condition (II.14). In view of (III.1), one then
sees that $x_N$ is non-linearly depended on the first (N-1) of $x_i$'s.
In the light of generalization
$$ K_{(Q_N)}(u_1,u_2,..,u_N) \equiv K(Q-N)  \eqno(III.8) $$
of (III.7), one can also see that all other Schur functions of degree
$ Q > N $ are degenerated for $A_{N-1}$ Lie algebras. This will also be
exemplified in the last section.

\vskip 3mm
\noindent {\bf{IV.\ WEYL CHARACTER FORMULA AND GENERALIZED SCHUR FUNCTIONS}}
\vskip 3mm

Throughout the sections IV and V, $\Lambda^+$ is an $A_{N-1}$ dominant weight
which corresponds, via (II.4), to a partition $(Q_k)$ of weight Q and length k.
We have two sets of indeterminates one of which is $\{ u_i , i=1,2,..N \}$
which come from the specialization (II.13) and the other one is
$ \{x_i , i=1,2,..N-1 \} $ which are created by the replacements (III.1).
The former set is constrained by the condition (II.14). It will be instructive
to emphasize here that Schur functions $S_{(Q_k)}(x_1,x_2,..,x_{N-1})$ depend
on  $x_i$'s while the right hand-side of Weyl formula will be obtained
in terms of indeterminates $u_i$'s. With this understanding, we will now show,
as similar as in (III.6), that for any other representation $R(\Lambda^+)$ the
Schur function $S_{(Q_k)}(x_1,x_2,..,x_{N-1})$ comes to be equivalent directly
to the right-hand side of Weyl character formula in the specialization (II.13).

The main object {\bf [8]} is
$$ A(\Lambda^+) \equiv \sum_{\omega} \ \epsilon(\omega) \
e^{\omega(\Lambda^+)} \eqno(IV.1) $$
where the sum is over $A_{N-1}$ Weyl group and hence $\omega(\Lambda^+)$
represents an action of Weyl reflection $\omega$ while $\epsilon(\omega)$
is its sign. Our objection here is on the explicit calculation of the sign
$\epsilon(\omega)$ which is known to be defined by
$$ \epsilon(\omega) = (-1)^{\ell(\omega)} \eqno(IV.2) $$
where $\ell(\omega)$ is the minimum number of simple reflections to obtain
Weyl reflection $\omega$. Instead, we can replace (IV.2) by
$$ \epsilon(\omega) = \epsilon_{q_{i_1},q_{i_2},..,q_{i_k}}  \eqno(IV.3) $$
for its action $\omega(\Lambda^+)$. As is mentioned above, recall here that
a reflection $\omega$ permutates the parameters $q_1,q_2,..,q_k$. The tensor
$ \epsilon_{q_{i_1},q_{i_2},..,q_{i_k}} $ is completely antisymmetric in its
indices while its numerical value is given on condition that
$$ \epsilon_{q_1,q_2,..,q_k} \equiv +1 \ \ , \ \
q_1 \geq q_2 \geq .. \geq q_k \ \ . $$

Weyl character formula now simply says that
$$ ChR(\Lambda^+) = {A(\rho+\Lambda^+) \over A(\rho)} \ \ \eqno(IV.3) $$
where $\rho \equiv \lambda_1 + \lambda_2 + .. + \lambda_{N-1} $ is $A_{N-1}$
{\bf Weyl vector} and the left-hand side of (IV.3) is determined by (II.9).
For finite Lie algebras, (IV.3) means that a miraculous factorization comes
out on the right hand side. By experience, no one could say that this is to be
seen so fast or so easy in practical applications for, especially, higher
dimensional representations of Lie algebras of higher rank. It would however
be greatly helpful to note, in the specialization (II.13), that
$$ A(\rho) = \prod_{j>i=1}^N \ (u_i-u_j)  \eqno(IV.4) $$
and $ A(\rho+\Lambda^+)$ can also be shown to be equivalent to determinant
of the following $N \times N$ matrix:
$$ \pmatrix{
u_1^{q_1-1+N}&u_1^{q_2-2+N}&\ldots&u_1^{q_N-N+N}\cr
u_2^{q_1-1+N}&u_2^{q_2-2+N}&\ldots&u_2^{q_N-N+N}\cr
\vdots&\vdots&\vdots&\vdots&\cr
u_N^{q_1-1+N}&u_N^{q_2-2+N}&\ldots&u_N^{q_N-N+N}\cr} \eqno(IV.5)   $$
where, for notational convenience, $ q_i \equiv 0 \ \ , \ \ i > k $.
(IV.4) is in fact nothing but the Vandermonde determinant and what is
miraculous here is the fact that $A(\rho+\Lambda^+)$ every time factorizes
into (IV.4) together with a polynomial which, for dominant $\Lambda^+$, can be
specified as a result of
$$ A(\rho+\Lambda^+) = A(\rho) \ S_{(Q_k)}(x_1,x_2,..,x_{N-1}) \ . \eqno(IV.6)  $$
In the case of $Q \ \lambda_1$, we already know from above that
$$ A(\rho+Q \ \lambda_1) = A(\rho) \ S_Q(x_1,x_2,..,x_{N-1}) \ . \eqno(IV.7) $$

This hence brings us back to the fact that {\bf Generalized Schur Functions
are in fact nothing but the right-hand side of Weyl character formula }
in the specialization (II.13).

\vskip 3mm
\noindent {\bf{V.\ $A_N$ MULTIPLICITY RULES}}
\vskip 3mm

In the light of above developments, the equivalence between (II.9) and (IV.3)
provides, for $A_{N-1}$, Weyl character formula in the form of
$$ ChR(\Lambda^+) = S_{(Q_k)}(x_1,x_2,..,x_{N-1}) \eqno(V.1) $$
and this hence allows us to calculate the multiplicities
$m_\Lambda^+(\lambda^+)$ without an explicit calculation of
$A(\rho+\Lambda^+)$, if one knows a way to calculate the Generalized Schur
Functions $ S_{(Q_k)}(x_1,x_2,..,x_{N-1}) $. It is so fortunate that this is
already provided by (I.3) because it serves us to express any
Generalized Schur Function in terms of the elementary ones in the form of
some kind of reduction formulas which we would like to call {\bf multiplicity
formulas} due to our way of using them in calculating weight multiplicities.
In view of (I.3), second order $A_{N-1}$ multiplicity rules, for instance,
are obtained, for k=2, in the form of
$$ S_{q_1,q_2} = S_{q_1} \  S_{q_2} - S_{q_1+1} \ S_{q_2-1} \eqno(V.2) $$
while third order multiplicity rules
$$ \eqalign{ S_{q_1,q_2,q_3} =
&+ S_{q_1+0} \  (S_{q_2+0} \ S_{q_3+0} - S_{q_2+1} \ S_{q_3-1}) \cr
&- S_{q_1+1} \  (S_{q_2-1} \ S_{q_3+0} - S_{q_2+1} \ S_{q_3-2}) \cr
&+ S_{q_1+2} \  (S_{q_2-1} \ S_{q_3-1} - S_{q_2+0} \ S_{q_3-2}) } \eqno(V.3) $$
are also obtained for k=3. The dependences on (N-1) indeterminates
$( x_1,x_2,..,x_{N-1})$ are suppressed above. It is now clear that we need an
explicit way to express Degenerated Schur Functions in order to use the
multiplicity rules as a powerful tool.

In applications of these multiplicity rules for $A_{N-1}$ Lie algebras,
the Elementary Schur Functions of degrees M=1,2,..N-1 are generic ones which
are already obtained by (I.1). As is explained above, all other ones are
Degenerated Schur Functions and one can easily investigate, in view of (III.7)
and (III.8), that they fulfill the following reduction rules:
$$ S_M = (-1)^N \ S_{M-N-1} - \sum_{k=1}^N \  S_k^{*} \ S_{M-k} \ \ ,
\ \ M \geq N \eqno(V.4) $$
where $S_k^{*}$ is obtained from $S_k$ under replacements
$x_i \rightarrow -x_i $ (i=1,2,..N-1.) .
In result, (V.1) gives us an overall equation in terms of several
monomials of (N-1) independent indeterminates $x_1, x_2,..,x_{N-1}$.
These monomials are homogeneous of degree $Q \equiv h(\Lambda^+)$ and
hence their numbers are just as the numbers of elements in
$Sub(Q \ \lambda_1)$. Since coefficients of these monomials are linearly
dependent on weight multiplicities, this means that we have, for each and
every element of $Sub(Q \ \lambda_1)$, a system of linear equations which
gives us unique solutions for weight multiplicities
$ m_{\Lambda^+}(\lambda^+) $ where $\lambda^+ \in Sub(Q \ \lambda_1)$.

Let us visualize the whole procedure in a simple example by studying,
for instance, all partitions $(7_k)$ \ (k=1,2,..6) in the notation of (IV.5) :

\hfill\eject

$$ \eqalign{
(7_1)_1 =& 7 + 0 + 0 + 0 + 0 + 0 \ \ , \ \ 7 \ \lambda_1 \cr
(7_2)_1 =& 6 + 1 + 0 + 0 + 0 + 0 \ \ , \ \ 5 \ \lambda_1 + \lambda_2 \cr
(7_2)_2 =& 5 + 2 + 0 + 0 + 0 + 0 \ \ , \ \ 3 \ \lambda_1 + 2 \ \lambda_2 \cr
(7_2)_3 =& 4 + 3 + 0 + 0 + 0 + 0 \ \ , \ \ \lambda_1 + 3 \ \lambda_2  \cr
(7_3)_1 =& 5 + 1 + 1 + 0 + 0 + 0 \ \ , \ \ 4 \ \lambda_1 + \lambda_3  \cr
(7_3)_2 =& 4 + 2 + 1 + 0 + 0 + 0 \ \ , \ \ 2 \ \lambda_1 + \lambda_2 + \lambda_3 \cr
(7_3)_3 =& 3 + 3 + 1 + 0 + 0 + 0 \ \ , \ \ 2 \ \lambda_2 + \lambda_3 \cr
(7_3)_4 =& 3 + 2 + 2 + 0 + 0 + 0 \ \ , \ \ \lambda_2 + 2 \ \lambda_3 \cr
(7_4)_1 =& 4 + 1 + 1 + 1 + 0 + 0 \ \ , \ \ 3 \ \lambda_2 + \lambda_4 \cr
(7_4)_2 =& 3 + 2 + 1 + 1 + 0 + 0 \ \ , \ \ \lambda_1 + \lambda_2 + \lambda_4 \cr
(7_4)_3 =& 2 + 2 + 2 + 1 + 0 + 0 \ \ , \ \ \lambda_3 + \lambda_4 \cr
(7_5)_1 =& 3 + 1 + 1 + 1 + 1 + 0 \ \ , \ \ 2 \ \lambda_1 + \lambda_5 \cr
(7_5)_2 =& 2 + 2 + 1 + 1 + 1 + 0 \ \ , \ \ \lambda_2 + \lambda_5 \cr
(7_6)_1 =& 2 + 1 + 1 + 1 + 1 + 1 \ \ , \ \ \lambda_1    } \eqno(V.5) $$
where an extra sub-index is added here to denote the partitions of the same
length and, for $A_5$, corresponding elements of $ Sub(7 \ \lambda_1) $  are
also shown on the right-hand side in view of (II.1-3). Weyl orbit characters
for these dominant weights can be obtained in terms of generators K(Q) by
the aid of the following reduction rules beyond third order:
$$ \eqalign{
&K_{(q_1,q_2,q_2,q_2)} = K(q_1) \ K_{(q_2,q_2,q_2)} - K_{(q_1+q_2,q_2,q_2)} \ \ , \ \ q_1 > q_2  \cr
&K_{(q_1,q_2,q_3,q_3)} = K(q_1) \ K_{(q_2,q_3,q_3)} - K_{(q_1+q_2,q_3,q_3)} -
K_{(q_1+q_3,q_2,q_3)}  \ \ , \ \ q_1 > q_2 > q_3  \cr
&K_{(q_1,q_1,q_1,q_2)} = 1/3 \ ( \ K(q_1) \ K_{(q_1,q_1,q_2)} -
K_{(q_1+q_1,q_1,q_2)} - K_{(q_1+q_2,q_1,q_1)} \ )  \ \ , \ \ q_1 > q_2 \cr
&K_{(q_1,q_2,q_2,q_2,q_2)} = K(q_1) \ K_{(q_2,q_2,q_2,q_2)} -
K_{(q_1+q_2,q_2,q_2,q_2)} \ \ , \ \ q_1 > q_2  \cr
&K_{(q_1,q_1,q_2,q_2,q_2)} = 1/2 \ ( \ K(q_1) \ K_{(q_1,q_2,q_2,q_2)} -
K_{(q_1+q_1,q_2,q_2,q_2)} - K_{(q_1+q_2,q_2,q_2,q_2)} \ )  \ \ , \ \ q_1 > q_2  }
\eqno(V.6) $$
All these class functions can therefore be assumed to be polynomials of
indeterminates $x_1,x_2,..,x_5$ because we have
$$ \eqalign{
x_6 =& -1 + {1 \over 720} x_1^6 - {1 \over 24} x_1^4 \ x_2 +
{1 \over 4} x_1^2 \ x_2^2 - \cr
&{1 \over 6} x_2^3 + {1 \over 6} x_1^3 \ x_3 -
x_1 \ x_2 \ x_3 + {1 \over 2} x_3^2 - {1 \over 2} x_1^2 \ x_4 +
x_2 \ x_4 + x_1 \ x_5  \cr
x_7 =& -x_1 + {1 \over 840} x_1^7 - {1 \over 30} x_1^5 \ x_2 +
{1 \over 6} x_1^3 \ x_2^2 + {1 \over 8} x_1^4 \ x_3 - \cr
&{1 \over 2} x_1^2 \ x_2 \ x_3 - {1 \over 2} x_2^2 \ x_3 -
{1 \over 3} x_1^3 \ x_4 + x_3 \ x_4 + {1 \over 2} x_1^2 \ x_5 +
x_2 \ x_5 } \eqno(V.7) $$
as a result of
$$ K_{(1,1,1,1,1,1)} \equiv 1 \ \ , \ \ K_{(2,1,1,1,1,1)} \equiv K(1) \ \ .
\eqno(V.8) $$
(V.7) is sufficient to obtain our first two Degenerated Schur Polynomials:
$$ \eqalign{
S_6 =& 1 + {1 \over 360} x_1^6 + {1 \over 2} x_1^2 \ x_2^2 +
{1 \over 3} x_1^3 \ x_3 + x_3^2 + 2 \ x_2 \ x_4 + 2 \ x_1 \ x_5 \cr
S_7 =& 2 \ x_1 + {1 \over 360} x_1^7 - {1 \over 15} x_1^5 \ x_2 +
{1 \over 2} x_1^3 \ x_2^2 + {1 \over 3}  x_1^4 \ x_3 - x_1^2 \ x_2 \ x_3 + \cr
&x_1 \ x_3^2 - {1 \over 3} x2 \ x_1^3 \ x_4 + 2 \ x_1 \ x_2 \ x_4 +
2 \ x_3 \ x_4 + 2 \ x_1^2 \ x_5 + 2 \ x_2 \ x_5 } \eqno(V.9)  $$
in addition to five generic ones $S_1,S_2,..,S_5$ which are known to be
obtained by the aid of (I.1). One can easily investigate that all these
are compatible with (V.4).

Let us now study Weyl character formula (V.1) in terms of above ingredients.
Its left-hand side is to be expressed in the form of an overall equation
which are expressed in terms of monomials
$$ x_1 , x_1^7 , x_1^5 \ x_2 , x_1^4 \ x_3 ,  x_1^3 \ x_2^2 , x_1^3 \ x_4 ,
x_1^2 \ x_5 , x_1^2 \ x_2 \ x_3 , x_1 \ x_2^3 , x_1 \ x_3^2 , x_1 \ x_2 \ x_4 ,
x_3 \ x_4 , x_1^2 \ x_5 , x_2 \ x_5 \ \ . \eqno(V.10) $$
with coefficients depending on weight multiplicities linearly. As is
emphasized above, note here especially that we have 14 monomials in (V.10)
while we also have 14 dominant weights in $Sub(7 \ \lambda_1)$. This
guarantees that the system of linear equations for multiplicities
$ m_\Lambda^+(\lambda^+) \ \ ( \lambda^+ \in Sub(7 \ \lambda_1))$ are always
consistent for each one of $R(\Lambda^+)$ with
$ \Lambda^+ \in Sub(7 \ \lambda_1)$. For the right-hand side of (V.1),
on the other hand, the multiplicity rules are sufficient in each one of these
cases. For instance, as being in line with (V.2),
$ S_{6,1} = S_6 \ S_1 - S_7 $ gives equivalently
$$ { 1 \over 15} \ (\ -15 \ x_1 + x_1^5 \ x_2 + 15 \ x_1^2 \ x_2 \ x_3 +
10 \ x_1^3 \ x_4 - 30 \ x_3 \ x_4 - 30 \ x_2 \ x_5  )  \eqno(V.11) $$
This is, in fact, nothing but a Degenerated Generalized Schur Function
and it provides the right-hand side of (V.1) by giving rise to solutions
$$ \eqalign{
&m_{5 \lambda_1 + \lambda_2}(7 \ \lambda_1) = 0  , \cr
&m_{5 \lambda_1 + \lambda_2}(5 \ \lambda_1 + \lambda_2) = 1 , \cr
&m_{5 \lambda_1 + \lambda_2}(3 \ \lambda_1 + 2 \ \lambda_2) = 1 \cr
&m_{5 \lambda_1 + \lambda_2}(\lambda_1 + 3 \ \lambda_2) = 1  , \cr
&m_{5 \lambda_1 + \lambda_2}(4 \ \lambda_1 + \lambda_3) = 2 , \cr
&m_{5 \lambda_1 + \lambda_2}(2 \ \lambda_1 + \lambda_2 + \lambda_3) = 2 \cr
&m_{5 \lambda_1 + \lambda_2}(2 \ \lambda_2 + \lambda_3) = 2 , \cr
&m_{5 \lambda_1 + \lambda_2}(\lambda_2 + 2 \ \lambda_3) = 2  , \cr
&m_{5 \lambda_1 + \lambda_2}(3 \ \lambda_2 + \lambda_4) = 3 \cr
&m_{5 \lambda_1 + \lambda_2}(\lambda_1 + \lambda_2 + \lambda_4) = 3 , \cr
&m_{5 \lambda_1 + \lambda_2}(\lambda_3 + \lambda_4) = 3  , \cr
&m_{5 \lambda_1 + \lambda_2}(2 \ \lambda_1 + \lambda_5) = 4 \cr
&m_{5 \lambda_1 + \lambda_2}(\lambda_2 + \lambda_5) = 4 , \cr
&m_{5 \lambda_1 + \lambda_2}(\lambda_1) = 3  } $$
for 1980-dimensional irreducible representation $R(5 \ \lambda_1 + \lambda_2)$
of $A_5$.

For further use, we give in the following all other Degenerated Generalized
Schur Functions participating in $Sub(7 \ \lambda_1)$:

$$ \eqalign{
S_{(5,2)} =& {1 \over 720} \ ( -720 \ x_1 + x_1^7 +
66 \ x_1^5 \ x_2 - 60 \ x_1^3 \ x_2^2 + 360 \ x_1 \ x_2^3 -
60 \ x_1^4 \ x_3 + 720 \ x_1^2 \ x_2 \ x_3 + \cr
&720 \ x_2^2 \ x_3 - 720 \ x_1 \ x_3^2 + 360 \ x_1^3 \ x_4 -
720 \ x_1 \ x_2 \ x_4 - 1080 \ x_1^2 \ x_5 + 720 \ x_2 \ x_5  )  \cr
S_{(4,3)} =& {1 \over 360} \ (
x_1^7 + 12 \ x_1^5 \ x_2 + 60 \ x_1^3 \ x_2^2 - 15 \ x_1^4 \ x_3 +
180 \ x_1^2 \ x_2 \ x_3 - 180 \ x_2^2 \ x_3 + \cr
&360 \ x_1 \ x_3^2 - 120 \ x_1^3 \ x_4 + 360 \ x_3 \ x_4 -
180 \ x_1^2 \ x_5 - 360 \ x_2 \ x_5  )  \cr
S_{(5,1,1)} =& {1 \over 240} \ (
240 \ x_1 + x_1^7 + 2 \ x_1^5 \ x_2 + 20 \ x_1^3 \ x_2^2 -
120 \ x_1 \ x_2^3 + 60 \ x_1^4 \ x_3 - 240 \ x_1^2 \ x_2 \ x_3 - \cr
&240 \ x_2^2 \ x_3 - 40 \ x_1^3 \ x_4 - 240 \ x_1 \ x_2 \ x_4 +
480 \ x_3 \ x_4 + 120 \ x_1^2 \ x_5 + 240 \ x_2 \ x_5   )  \cr
S_{(4,2,1)} =& {1 \over 120}  (
120 \ x_1 + x_1^7 + 20 \ x_1^3 \ x_2^2 + 15 \ x_1^4 \ x_3 -
180 \ x_1^2 \ x_2 \ x_3 - \cr
&60 \ x_2^2 \ x_3 - 80 \ x_1^3 \ x_4 +
240 \ x_1 \ x_2 \ x_4 - 120 \ x_3 \ x_4 + 120 \ x_1^2 \ x_5    )  \cr
S_{(3,3,1)} =& {1 \over 240} \ (
x_1^7 + 2 \ x_1^5 \ x_2 + 20 \ x_1^3 \ x_2^2 - 120 \ x_1 \ x_2^3 -
30 \ x_1^4 \ x_3 + 120 \ x_1^2 \ x_2 \ x_3 +  \cr
&120 \ x_2^2 \ x_3 - 40 \ x_1^3 \ x_4 - 240 \ x_1 \ x_2 \ x_4 -
240 \ x_3 \ x_4 + 120 \ x_1^2 \ x_5 + 240 \ x_2 \ x_5      )   } $$

$$ \eqalign{
S_{(3,2,2)} =& {1 \over 240} \ (
x_1^7 - 2 \ x_1^5 \ x_2 + 20 \ x_1^3 \ x_2^2 + 120 \ x_1 \ x_2^3 -
30 \ x_1^4 \ x_3 - 120 \ x_1^2 \ x_2 \ x_3 +  \cr
&120 \ x_2^2 \ x_3 + 40 \ x_1^3 \ x_4 - 240 \ x_1 \ x_2 \ x_4 +
240 \ x_3 \ x_4 + 120 \ x_1^2 \ x_5 - 240 \ x_2 \ x_5     )   \cr
S_{(4,1,1,1)} =& {1 \over 360} \ (
-360 \ x_1 + x_1^7 + 12 \ x_1^5 \ x_2 - 180 \ x_1^3 \ x_2^2 -
15 \ x_1^4 \ x_3 + 180 \ x_1^2 \ x_2 \ x_3 + \cr
&540 \ x_2^2 \ x_3 + 360 \ x_1 \ x_3^2 + 120 \ x_1^3 \ x_4 -
360 \ x_3 \ x_4 - 180 \ x_1^2 \ x_5 - 360 \ x_2 \ x_5     )   \cr
S_{(3,2,1,1)} =& {1 \over 360} \ (
-360 \ x_1 + 2 \ x_1^7 - 120 \ x_1^3 \ x_2^2 - 75 \ x_1^4 \ x_3 +
540 \ x_1^2 \ x_2 \ x_3 -  \cr
&180 \ x_2^2 \ x_3 - 360 \ x_1 \ x_3^2 +
240 \ x_1^3 \ x_4 + 360 \ x_3 \ x_4 - 360 \ x_1^2 \ x_5     )   \cr
S_{(2,2,2,1)} =& {1 \over 360} \ (
x_1^7 - 12 \ x_1^5 \ x_2 + 60 \ x_1^3 \ x_2^2 - 15 \ x_1^4 \ x_3 -
180 \ x_1^2 \ x_2 \ x_3 - 180 \ x_2^2 \ x_3 +  \cr
&360 \ x_1 \ x_3^2 + 120 \ x_1^3 \ x_4 - 360 \ x_3 \ x_4 -
180 \ x_1^2 \ x_5 + 360 \ x_2 \ x_5)/360   \cr
S_{(3,1,1,1,1)} =& { 1 \over 240} \ (
240 \ x_1 + x_1^7 - 18 \ x_1^5 \ x_2 + 20 \ x_1^3 \ x_2^2 +
120 \ x_1 \ x_2^3 + 60 \ x_1^4 \ x_3 - \cr
&240 \ x_2^2 \ x_3 - 120 \ x_1^3 \ x_4 -
240 \ x_1 \ x_2 \ x_4 + 120 \ x_1^2 \ x_5 + 240 \ x_2 \ x_5  ) \cr
S_{(2,2,1,1,1)} =& { 1 \over 240 } \ (
240 \ x_1 + x_1^7 - 22 \ x_1^5 \ x_2 + 100 \ x_1^3 \ x_2^2 -
120 \ x_1 \ x_2^3 + 60 \ x_1^4 \ x_3 - \cr
&240 \ x_1^2 \ x_2 \ x_3 + 240 \ x_2^2 \ x_3 - 120 \ x_1^3 \ x_4 +
240 \ x_1 \ x_2 \ x_4 + 120 \ x_1^2 \ x_5 - 240 \ x_2 \ x_5    )  } $$

We finally add that all these calculations can be performed by the aid of
some simple computer programs written, for instance, in the language of
Mathematica {\bf [14]}. They can be provided by e-mail from the address
noted in the first page of this paper.

\vskip3mm
\noindent{\bf {REFERENCES}}
\vskip3mm
\leftline{[1] H.Freudenthal, Indag. Math. 16 (1964) 369-376}

\leftline{[2] G.Racah, Group Theoretical Concepts and Methods in Elementary Particle Physics,}
\leftline{\ \ \ ed. F.Gursey, N.Y., Gordon and Breach (1964) }

\leftline{[3] B.Kostant, Trans.Am.Math.Soc., 93 (1959) 53-73}

\leftline{[4] R.C.King and S.P.O.Plunkett, J.Phys.A:Math.Gen., 9 (1976) 863-887}
\leftline{ \ \ \ R.C.King and A.H.A Al-Qubanchi, J.Phys.A:Math.Gen., 14 (1981) 51-75}
\leftline{ \ \ \ K.Koike, Jour.Algebra, 107 (1987) 512-533}
\leftline{ \ \ \ R.V.Moody and J.Patera, Bull.Am.Math.Soc., 7 (1982) 237-242}

\leftline{[5] H.Weyl, The Classical Groups, N.J. Princeton Univ. Press (1946)}

\leftline{[6] I.B.Frenkel, Representations of Kac-Moody Algebras and Dual Resonance Models,}
\leftline{ \ \ \ Lectures in Applied Mathematics 21 (1985) 325-353}

\leftline{[7] M.Bremner, R.V.Moody and J.Patera, Tables of Dominant Weight Multiplicities for Representations}
\leftline{ \ \ \ of Simple Lie Algebras, N.Y., M.Dekker (1985)}

\leftline{[8] V.G.Kac, Infinite Dimensional Lie Algebras, N.Y., Cambridge Univ. Press (1990)}

\leftline{[9] S.Kass, R. Moody, J.Patera and R.Slansky, Affine Kac-Moody Algebras, Weight Multiplicities and }
\leftline{ \ \ \ Branching Rules, Berkeley, Univ. California Press (1990)}

\leftline{[10] V.G.Kac and S.Peterson, Adv. Math., 53 (1984) 125-264}

\leftline{[11] V.G.Kac and A.K.Raina, Bombay Lectures on Highest Weight Representations,}
\leftline{\ \ \ \ \ World Sci., Singapore (1987)}

\leftline{[12] H.R.Karadayi, Anatomy of Grand Unifying Groups I and II ,}
\leftline{ \ \ \ \ ICTP preprints(unpublished) IC/81/213 and 224}
\leftline{ \ \ \ \  H.R.Karadayi and M.Gungormez, Jour.Math.Phys., 38 (1997) 5991-6007}

\leftline{[13] J.E.Humphreys, Introduction to Lie Algebras and Representation Theory, N.Y., Springer-Verlag (1972)}

\leftline{[14] S. Wolfram, Mathematica$^{TM}$, Addison-Wesley (1990) }

\end